\documentstyle[aps,prl,epsf,preprint,tighten,floats,amssymb,amsfonts]{revtex}

\begin{document}
\draft
\title{Nonuniversality in quantum wires with off-diagonal disorder:\\
a geometric point of view}

\author{P. W. Brouwer,$^a$ C. Mudry,$^a$ and A.\ Furusaki$^b$}
\address{$^a$Lyman Laboratory of Physics, Harvard University, Cambridge MA 02138\\$^b$Yukawa Institute for Theoretical Physics, Kyoto University,
Kyoto 606-8502, Japan}
\date{\today}
\maketitle 

\begin{abstract}
It is shown that, in the scaling regime, transport properties of
quantum wires with off-diagonal disorder are described by a family of
scaling equations that depend on two parameters: the mean free path and
an additional continuous parameter. The existing scaling equation for
quantum wires with off-diagonal disorder [Brouwer et al.,
Phys.\ Rev.\ Lett.\ {\bf 81}, 862 (1998)] is a special point in this
family. Both parameters depend on
the details of the microscopic model. Since there are two parameters
involved, instead of only one, localization in a wire with
off-diagonal disorder is not universal. We take a geometric point of
view and show that this nonuniversality follows from the fact that the
group of transfer matrices is not semi-simple. Our results are
illustrated with numerical simulations for a tight-binding model with
random hopping amplitudes. \medskip\\
\end{abstract}

Universality is a key concept in any approach to study localization in
disordered systems. In the scaling theory of localization, it is
commonly believed that the statistical distributions of the
conductance, or of energy levels and wave functions, are entirely
determined by the fundamental symmetries and the dimensionality of the
sample \cite{LeeReview}. Once symmetry and dimensionality are taken
into account, all microscopic details of a sample can be represented by
a single length scale $\ell$, the ``mean free path'', such that on
length scales $L$ larger than $\ell$, the sample is completely
characterized by the ratio $L/\ell$. The concept of universality is the
cornerstone of various field-theoretic, diagrammatic, and random-matrix
approaches to localization
\cite{LeeReview,StoneReview,AltshulerReview,EfetovBook}.

For a quasi-one dimensional geometry, i.e., for quantum wires, and for
weak disorder (mean free path $\ell$ is much larger than the Fermi
wavelength $\lambda$) the statistical distribution of the conductance
can be described by the transfer matrix approach of Dorokhov
\cite{Dorokhov}, and Mello, Pereyra, and Kumar \cite{MPK} (DMPK). In
this approach, the transport properties of the quantum wire are
described in terms of a scaling equation for its transmission
eigenvalues, the so-called DMPK equation \cite{StoneReview}.
H\"uffmann \cite{Hueffmann} and Caselle \cite{Caselle} have provided
the DMPK equation with a geometric foundation by reformulating it in
terms of a Brownian motion of the transfer matrix on a symmetric space,
a certain curved manifold from the theory of Lie groups and Lie
algebras \cite{Helgason}. The existence of a unique natural
mathematical framework to describe Brownian motion on symmetric spaces
provides the geometric counterpart of the observed universality of
the localization properties of disordered quantum wires.

In a recent work, two of the authors, together with Simons and Altland
\cite{BMSA}, have proposed an extension of the scaling approach of DMPK
to quantum wires with off-diagonal disorder (e.g.\ a lattice model with
random hopping amplitudes and no on-site disorder).  At the band
center $\varepsilon=0$ such a system has an extra chiral or
sublattice symmetry that is not present in the standard case of a
wire with diagonal (potential) disorder.  Therefore they belong to a
different symmetry class, which is referred to as the {\em chiral}
symmetry class. Chiral symmetry also plays an important role for
two-dimensional Dirac fermions in a random vector potential
\cite{Ludwig,Nersesyan,Mudry}, the lattice random flux model
\cite{Lee,Miller96,Furusaki}, non-Hermitian quantum mechanics
\cite{Sommers,HatanoNelson}, supersymmetric quantum mechanics
\cite{Comtet}, diffusion in a random medium \cite{Bouchaud}, and in
certain problems in QCD \cite{Verbaarschot}.

In Ref.\ \onlinecite{BMSA}, the extension of the DMPK equation to the
chiral symmetry class was derived from a simple microscopic model.
Here we discuss its geometric origin. This proves to be an exercise with
implications that reach far beyond the construction of a
mere chiral parallel to the geometric foundation of the DMPK
equation of Refs.\ \onlinecite{Hueffmann,Caselle}:
Upon inspection of the geometric structure underlying the ``chiral DMPK
equation'', we find that the localization properties of a disordered
wire with off-diagonal disorder are {\em not} universal; the geometric
approach allows for a one-parameter family of scaling equations for the
transmission eigenvalues. The scaling equation that was originally
found in Ref.\ \onlinecite{BMSA} is a special point in this family.
Below we discuss the geometric approach in more detail, identify the
origin of this non-universality, and illustrate the results with
numerical simulations of different microscopic models.

\begin{figure}
\epsfxsize=0.7\hsize
\hspace{0.15\hsize}
\epsffile{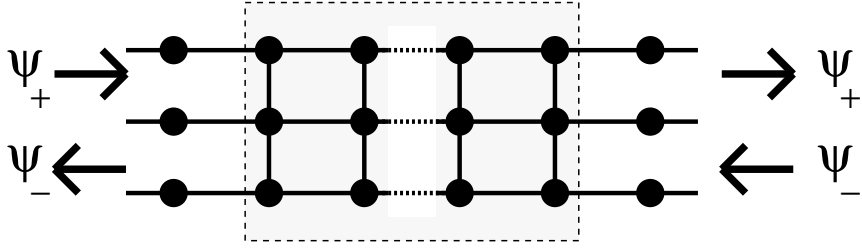}\bigskip

\caption{\label{fig:0} Lattice model for a disordered sample consisting
of $N=3$ coupled chains (dotted region) attached to two ideal leads. In
the two leads, the wave function is represented by two $N$-component vectors
of amplitudes for the left and right moving waves respectively. In
the disordered region, the Hamiltonian contains a random on-site potential.}
\end{figure}

We start with a brief summary of the ideas that lead to the standard
DMPK equation and its geometric interpretation. The cornerstone of this
approach are the symmetry properties of the transfer matrix $M$ of a
disordered wire. For definiteness, we focus on the lattice Anderson
model \cite{AltshulerReview} in a geometry of $N$ coupled chains, see
Fig.\ \ref{fig:0}. In this model, the Hamiltonian consists of nearest
neighbor hopping terms and a random on-site potential. We restrict our
attention to spinless particles. The wire consists of a disordered
region and of two ideal leads consisting of $N$ (uncoupled) chains. In
the leads, the wave function is represented by an $N$-component vector
$\psi_{\pm}$ for the amplitudes of left ($+$) and right ($-$) moving
waves. The wave functions to the left and right sides of the
disordered sample are related by the $2N \times 2N$ transfer matrix $M$
\cite{StoneReview},
\begin{equation}
  { \psi_{+} \choose \psi_{-}}_{\rm right} = M
  { \psi_{+} \choose \psi_{-}}_{\rm left}.
\end{equation}
In this basis of left and right movers, flux conservation implies that
the transfer matrix $M$ obeys
\begin{equation}
  M^{\dagger} \Sigma_3 M = \Sigma_3, \label{eq:flux}
\end{equation}
where $\Sigma_3 = \sigma_3 \otimes \openone_N$, $\sigma_3$ being the
Pauli matrix and $\openone_N$ the $N \times N$ unit matrix. We 
distinguish between the presence and absence of time-reversal symmetry,
labeled by the symmetry parameter $\beta=1,2$, respectively. In the
presence of time-reversal symmetry, $M$ further satisfies
\begin{equation}
  \Sigma_1 M^{*} \Sigma_1 = M,
\end{equation}
where $\Sigma_1 = \sigma_1 \otimes \openone_N$. The transfer matrix $M$
can be parametrized as \cite{StoneReview,Hua}
\begin{equation}
  M = \left( \begin{array}{cc} U & 0 \\ 0 & U' \end{array} \right)
\left( \begin{array}{cc} \cosh x & \sinh x \\ \sinh x & \cosh x \end{array} \right)
\left( \begin{array}{cc} V & 0 \\ 0 & V' \end{array} \right),
  \label{eq:polar}
\end{equation}
where $U$, $U'$, $V$, and $V'$ are unitary matrices, and $x$ is a
diagonal matrix with diagonal elements $x_j$. For $\beta=1$, $V' =
V^{*}$ and $U' = U^{*}$. The unitary matrices $U$, $U'$, $V$, and $V'$ 
serve as ``angular coordinates'' for $M$, the parameters $x_j$ serve
as ``radial coordinates'' [the eigenvalues of $M M^{\dagger}$ are
$\exp(\pm 2 x_j)$]. The radial coordinates $x_j$ are related to the
transmission eigenvalues $T_j$ by $T_j = 1/\cosh^2 x_j$. They determine
the dimensionless conductance $g$ through the Landauer formula,
\begin{equation}
  g = \sum_{j=1}^{N} T_j = \sum_{j=1}^{N} {1 \over \cosh^2 x_j}.
\end{equation}

In the original derivations of the DMPK equation \cite{Dorokhov,MPK},
the wire is divided into thin slices, and the transfer matrix $M$ of
the entire wire is found by multiplication of the transfer matrices of
the individual slices. For weak disorder, the parameters $x_j$ of the transfer matrix
undergo only small changes $x_j \to x_j + \delta x_j$ upon each such
multiplication. One can view this process as a ``Brownian motion'' for
the parameters $x_j$, the length $L$ of the wire serving as a
fictitious time. With the choice of a maximum information entropy
distribution for the transfer matrix of the slice \cite{MPK}, one can
write down the corresponding Fokker-Planck equation, which has the form
\begin{eqnarray}
  {\partial P \over \partial L} &=& D
  \sum_{j} {\partial \over \partial x_j} J 
  {\partial \over \partial x_j} J^{-1} P, \nonumber \\
  J &=& \prod_{j} \sinh (2 x_j) \prod_{j<k} \sinh^{\beta}(x_j - x_k)
\sinh^{\beta}(x_j + x_k). \label{eq:DMPK}
\end{eqnarray}
The proportionality constant $D$ is determined by the details
of the microscopic model.\footnote{
        In the DMPK equation, $D^{-1}=2(\beta N + 2 - \beta) \ell$, 
        where $\ell$ is the mean free path, see e.g.\
        Ref.\ \onlinecite{StoneReview,Hueffmann}.
}

H\"uffmann pointed out that there is a beautiful geometric structure
underlying the scaling equation (\ref{eq:DMPK}). He observed that
Eqs.\ (\ref{eq:flux})--(\ref{eq:polar}) express that the transfer
matrix $M$ is a member of a Lie group $G_{\beta}$, with\footnote{
	$U(N,N)$ is the group of complex matrices $M$ with $M^{\dagger}
	\Sigma_3 M = \Sigma_3$; $SU(N,N)$ is its subgroup of matrices
	with unit determinant; ${Sp}(N,{\Bbb R})$ is the group of real
	$2N \times 2N$ matrices $M$ that obey $M^T \Sigma_2 M =
	\Sigma_2$, and $Z_2 = \{\pm 1\}$.  To see that $G_1 \simeq Z_2
	\times {Sp}\,(N,{\Bbb R})$, see
	e.g.\ Ref.\ \onlinecite{StoneReview}.
}
\begin{equation}
  G_{1} \simeq Z_2 \times Sp(N,{\Bbb R}), \ \ \ \
  G_{2} \simeq U(N,N) \simeq U(1) \times SU(N,N),
\end{equation}
and proposed to describe the $L$ evolution of $M$ as a ``Brownian
motion'' on the manifold $G_{\beta}$ \cite{Hueffmann}. 
For a rigorous formulation of
this Brownian motion process, two further steps have to be taken.
We would like to repeat them here,
as we need to reconsider them when we deal with the case of
a quantum wire with off-diagonal disorder.
\begin{enumerate}
\item  The Lie groups $G_{\beta}$ are not semi-simple: they
       are the direct product of the two components $Z_2$ and
       ${Sp}\,(N,{\Bbb R})$, or $U(1)$ and $SU(N,N)$, for $\beta=1$ or $2$
       respectively. For the product $M M^{\dagger}$, and hence for the
       radial coordinates $x_j$, only the non-compact components
       ${Sp}\,(N,{\Bbb R})$ and $SU(N,N)$ are relevant, so that we may
       restrict our attention to the semi-simple Lie groups
       ${Sp}\,(N,{\Bbb R})$ and $SU(N,N)$. The remaining components $Z_2$
       and $U(1)$ correspond to the sign or phase of $\det M$ and do
       not affect the $x_j$.
\item  The algebraic structure on the semi-simple Lie groups 
       ${Sp}\,(N,{\Bbb R})$ and $SU(N,N)$ gives rise to a natural metric
       \cite{Helgason}.  However, since this natural metric is not
       positive definite, it cannot be used to define a Brownian motion
       process. The problem is solved by dividing out the maximal
       compact subgroups $U(N)$ or $S(U(N) \times U(N))$ for $\beta=1$
       or $2$, respectively. The resulting coset spaces $S_1 =
       Sp(N,{\Bbb R})/U(N)$ and $S_2 = SU(N,N)/S(U(N) \times U(N))$ are
       called symmetric spaces and have a natural positive definite
       metric \cite{Helgason}.  In the parametrization
       (\ref{eq:polar}), the procedure of dividing out the subgroups
       $U(N)$ or $S(U(N) \times U(N))$ corresponds to the
       identification of all transfer matrices $M$, $M'$ for which the
       product $M^{-1} M'$ is of the form
       \begin{equation}
         M^{-1} M' = \left( \begin{array}{cc} V & 0 \\ 
                            0 & V' \end{array} \right), 
       \end{equation} 
       where $V$ and $V'$ are arbitrary unitary matrices ($V' = V^{*}$
       for $\beta=1$ and $\det V' V = 1$ for $\beta=2$).
       In shifting to the symmetric spaces $S_{\beta}$
       no information on the radial coordinates $x_j$ is lost, because
       they are well-defined in each equivalence class.

       The symmetric spaces $S_{\beta}$ admit a spherical coordinate
       system whose radial coordinates $x_1,\ldots,x_N$ are equal to
       the radial coordinates $x_1,\ldots,x_N$ of the transfer matrix
       $M$, cf.\ Eq.\ (\ref{eq:polar}). The Brownian motion of the
       $x_j$ is described by the radial part of the Laplace-Beltrami
       operator on $S_{\beta}$. This radial part is known from the
       literature \cite{Hueffmann,Caselle,Helgason}, and is found to be 
       identical to the differential operator in Eq.\ (\ref{eq:DMPK}). 
       As a result, we find that
       their probability distribution $P(x_1,\ldots,x_N;L)$ satisfies
       the Fokker-Planck equation (\ref{eq:DMPK}), where the constant
       $D$ is the diffusion constant on $S_{\beta}$. The appearance of
       a single diffusion constant $D$ signifies the universality of
       the localization properties in disordered quantum wires.
\end{enumerate}

Let us now consider a quantum wire with off-diagonal disorder (random
hopping), at the band center $\varepsilon = 0$. We consider the same
geometry as in Fig.\ \ref{fig:0}, the only difference being that now
the randomness is in the hopping amplitudes between neighboring lattice
site, the on-site potential being zero everywhere. On a bipartite
lattice (which is the case we consider here, cf.\ Fig.\ \ref{fig:0}), and
with only off-diagonal (hopping) randomness, localization is different
from the standard case described above because of the existence of an
additional symmetry, known as a sublattice or chiral symmetry
\cite{BMSA},
\begin{equation}
       \Sigma_1 M \Sigma_1 = M. \label{eq:chiral}
\end{equation}
Hence the unitary matrices in the
parametrization (\ref{eq:polar}) satisfy $U = U'$ and $V = V'$.
Let us now consider the extension of the DMPK equation (\ref{eq:DMPK})
that includes the chiral symmetry (\ref{eq:chiral}), following
H\"uffmann's geometric approach.

The Lie groups $G^{\rm ch}_{\beta}$ of transfer matrices $M$ that
obey the chiral symmetry (\ref{eq:chiral}) are\footnote{
        $GL(N,{\Bbb R})$ and $GL(N,{\Bbb C})$ are the
        multiplicative groups of $N \times N$ matrices with real and complex
        elements, respectively; $SL(N,{\Bbb R})$ and $SL(N,{\Bbb C})$ are their
        subgroups of matrices with unit determinant; ${\Bbb R}^{+}$ is the
        multiplicative group of the positive real numbers. The origin
        of these transfer matrix groups is explained below
        Eq.\ (\ref{eq:solution}).
}
\begin{eqnarray*}
  G^{\rm ch}_{1} &\simeq& GL(N,{\Bbb R}) \simeq
  Z_2 \times {\Bbb R}^{+} \times SL(N,{\Bbb R}),\\
  G^{\rm ch}_{2} &\simeq& GL(N,{\Bbb C}) \simeq U(1) \times {\Bbb R}^{+} 
    \times SL(N,{\Bbb C}).
\end{eqnarray*}
The construction of a Brownian motion process for $M$ on
$G_{\beta}^{\rm ch}$ proceeds with the same two steps as in the standard
case of diagonal disorder. However, there is an important difference.
Unlike the transfer matrix groups $G_{\beta}$ discussed above, the
transfer matrix groups $G_{\beta}^{\rm ch}$ for off-diagonal disorder
contain {\em two} non-compact factors, ${\Bbb R}^{+}$ and $SL(N,{\Bbb R})$,
or ${\Bbb R}^{+}$ and $SL(N,{\Bbb C})$, for $\beta=1$ or $2$, respectively.
(This difference was noted by Zirnbauer in a field-theoretic context
\cite{Zirnbauer}.) Both of them determine the radial coordinates
$x_j$:  The factor ${\Bbb R}^{+}$ describes the position of the average
$\bar x = (x_1 + \ldots + x_N)/N$, while the special linear group
$SL(N)$ is connected to the relative positions of the $x_j$. One
therefore has to consider two different Brownian motion processes: One
on ${\Bbb R}^{+}$, to describe the $L$ evolution of $\bar x$, and one on the
symmetric space $S_{1}^{\rm ch} = SL(N,{\Bbb R})/SO(N)$ or $S_2^{\rm
ch} = SL(N,{\Bbb C})/SU(N)$, to describe the $L$ evolution of the
differences of the radial coordinates $x_j$.  [The symmetric spaces
$S_{\beta}^{\rm ch}$ are obtained after dividing out the maximal compact
subgroup of $SL(N,{\Bbb R})$ or $SL(N,{\Bbb C})$, as in the standard 
case.] Since these two
Brownian motion processes have two different and a priori unrelated
diffusion constants, one needs their ratio as an extra parameter to
characterize the distribution of the transfer matrix. This is the
absence of universality in quantum wires with off-diagonal disorder
that we announced in the introduction.

In this most general case, there is a one-parameter family of
Fokker-Planck equations for the distribution $P(x_1,\ldots,x_N;L)$ of
the radial coordinates $x_j$. This family of Fokker-Planck equations is
constructed in two steps.  First, Brownian motion on ${\Bbb R}^{+}$
results in a simple diffusion equation for the distribution $P_{\bar
x}(\bar x;L)$ of the average $\bar x$,
\begin{equation}
  {\partial P_{\bar x} \over \partial L} =
  {D_R \over N} {\partial^2 \over \partial \bar x^2} P_{\bar x}.
  \label{eq:xbar}
\end{equation}
Here $D_R/N$ is the diffusion coefficient for the Brownian motion on
${\Bbb R}^{+}$. (We chose to write the diffusion coefficient as $D_R/N$ for
later convenience; it is the diffusion coefficient that one would
obtain for the average $\bar x$ if the $N$ variables $x_j$
would diffuse independently with diffusion coefficient $D_R$.)
Second, Brownian motion on $S^{\rm ch}_{\beta}$ is described in terms of radial
coordinates $y_j$, $j=1,\ldots,N$, with the constraint $y_1 + \ldots +
y_N = 0$ \cite{Caselle,Helgason}. They correspond to the radial
coordinates $x_j$ of the transfer matrix $M$ via $y_j = x_j - \bar x$.
Using the explicit form for the Laplace-Beltrami operator on the
symmetric spaces $S_{\beta}^{\rm ch}$, the Fokker-Planck equation for
the probability distribution $P_y(y_1,\ldots,y_N;L)$ of the $y_j$ reads
\begin{equation}
  {\partial P_y \over \partial L} =
  D_S \sum_{j=1}^{N} {\partial \over \partial y_j} J 
  {\partial \over \partial y_j} J^{-1} P_y,\ \ \ \
  J = \prod_{j < k} \sinh^{\beta}(y_j - y_k),
\end{equation}
where we have to restrict to the subspace $y_1 + \ldots + y_N = 0$.
Hence, for the distribution $P(x_1,\ldots,x_N;L)$ of the radial
coordinates $x_j = \bar x + y_j$ of the transfer matrix $M$ we find 
the Fokker-Planck equations
\begin{equation}
  {\partial P \over \partial L} = 
  D_{\rm ch} \sum_{j,k=1}^{N} {\partial \over \partial x_j} 
    \left(\delta_{jk} - {1-\eta \over N} \right) J
    {\partial \over \partial x_k} J^{-1} P, \ \  \ \
  J = \prod_{j < k} \sinh^{\beta}(x_j - x_k), \label{eq:DMPKchiral}
\end{equation}
where $D_{\rm ch} = D_S$ and $\eta = D_R/D_S$. Equation
(\ref{eq:DMPKchiral}) contains the full one-parameter family of
scaling equations that describes localization at the band center
$\varepsilon = 0$ in quantum wires with off-diagonal disorder.  The
case $\eta=1$ (i.e., equal diffusion constants $D_R$ and $D_S$)
corresponds to the equation that was derived before in
Ref.\ \onlinecite{BMSA}.

A solution of the scaling equation (\ref{eq:DMPKchiral}) in the
localized regime $L \gg D_{\rm ch}$ can be obtained by standard methods
\cite{StoneReview}. The distribution of the radial coordinates is
Gaussian, with average and variance given by
\begin{equation}
  \langle x_j \rangle = (N+1-2j) \beta L D_{\rm ch},\ \ \ \
  \langle x_k x_j \rangle - \langle x_j \rangle \langle x_k \rangle =
    2 \left(\delta_{jk} - {1-\eta \over N} \right)L D_{\rm ch}.
\end{equation}
For even $N$, the distribution of the conductance $g$ is log-normal, with
\begin{eqnarray}
  \langle \ln g \rangle &=& \nonumber
    -\beta s + 2 \left[{1 \over \pi} \left(1 - 2 {1-\eta \over N}\right)\right]^{1/2} s^{1/2} + 
    {\cal O}(1), \\
  \mbox{var}\, \ln g &=& 2 \left[ 1 + \left(1-{2 \over \pi} \right)\left(1-2{1-\eta \over N} \right) \right] s
  + {\cal O}(1), \label{eq:lnG}
\end{eqnarray}
where $s = 2LD_{\rm ch}$. For odd $N$, the fluctuations of $\ln G$ are
of the same order as the average,
\begin{eqnarray}
  \langle \ln g \rangle &=& - 2 \left[{2 \over \pi}
    \left(1-{1-\eta \over N}\right)\right]^{1/2} s^{1/2} +
    {\cal O}(1), \nonumber \\ \label{eq:lnGodd}
  \mbox{var}\, \ln g &=& 4 \left(1-{2 \over \pi} \right) \left(1 - {1-\eta \over N} \right) s + {\cal O}(1).
\end{eqnarray}
For comparison, with diagonal disorder, $\langle \ln g \rangle =
-4DL$ and $\mbox{var}\, \ln g = 8DL$ \cite{StoneReview}. Although
for even $N$ the conductance distribution is log-normal both for
diagonal and for off-diagonal disorder, there are some important
differences: 
The presence of a term $\propto L^{1/2}$ in $\langle \ln g \rangle$, the
$\beta$-dependence of $\langle \ln g \rangle$, and the absence of
universal fluctuations of $\ln g$ (they depend on $\eta$) are special
for off-diagonal disorder and do not appear in the standard case
of diagonal disorder \cite{StoneReview}.

\begin{figure}
\epsfxsize=0.95\hsize
\hspace{0.04\hsize}
\epsffile{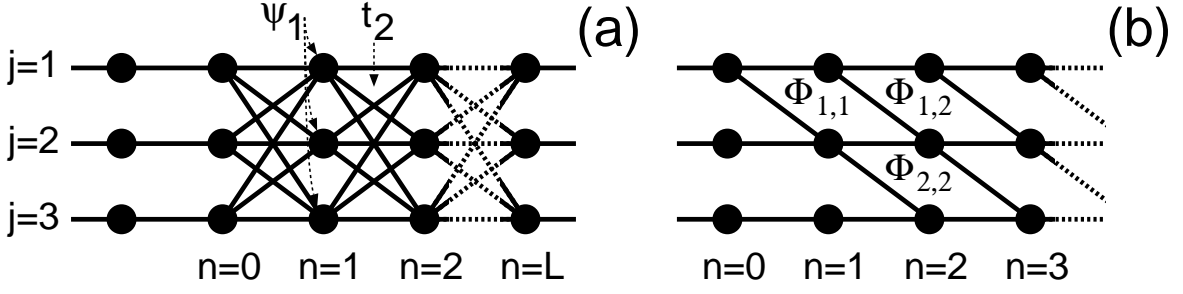}
\caption{\label{fig:1} 
  (a) Most general random hopping chain corresponding to the
  Schr\"odinger Equation (\protect\ref{eq:Schrod}) for $N=3$.
  (b) Random flux model: A square lattice where each plaquette is
  threaded by a random flux $\Phi$.}
\end{figure}

What is $\eta$ for a particular microscopic model? Although we cannot
answer this question in general, we can discuss some examples of
microscopic models of quantum wires with hopping disorder and compare
theoretical predictions from Eq.\ (\ref{eq:DMPKchiral}) with numerical
simulations.
We start from the Schr\"odinger equation for a set of
coupled chains with random hopping, which in general can be written as
\begin{equation}
  \varepsilon \psi_n = -t_{n}^{\vphantom{\dagger}} \psi_{n+1}
   - t_{n-1}^{\dagger} \psi_{n-1}. \label{eq:Schrod}
\end{equation}
Here $n$ labels the position along the wire (in units of the lattice
spacing), $\psi_n$ is an $N$-component wavevector, and $t_n$ is an $N
\times N$ hopping matrix, see Fig.\ \ref{fig:1}a. (This is a more
general formulation than the one of Fig.\ \ref{fig:0}. Note that the
lattice of Fig.\ \ref{fig:0} is in this class.) The hopping matrices
$t_n$ are real (complex) for $\beta=1$ ($2$). We connect the disordered
region of length $L$ to two ideal leads for $n < 0$ and $n > L$,
characterized by $t_n = \openone$. For zero energy it is possible to
solve for the transfer matrix explicitly,
\begin{equation}
  M(L) = {1 \over 2} \left( \begin{array}{cc} m^{\dagger} + m^{-1} & m^{\dagger} - m^{-1} \\
    m^{\dagger} - m^{-1} & m^{\dagger} + m^{-1} \end{array} \right),\ \ \ \
  m = \prod_{n=1}^{L/2} (t_{2n-1} t_{2n}^{\dagger-1}).
  \label{eq:solution}
\end{equation}
(For convenience, we assumed that $L$ is even; recall that we use
a basis of left and right movers.)
The product $m^{\dagger} m$ has eigenvalues $\exp(2 x_j)$,
$j=1,\ldots,N$; the product $m^{-1} m^{-1\dagger}$ has eigenvalues
$\exp(-2 x_j)$.  Being an arbitrary $N \times N$ matrix with real or
complex elements, the matrix $m$ is an element of the linear group
$GL(N,{\Bbb R})$ or $GL(N,{\Bbb C})$, and, taking into account the considerations
outlined above, the $L$ dependence of the transfer matrix $M$ can be
described by the random trajectory of the matrix $m$ in $GL(N,{\Bbb R})$ 
or $GL(N,{\Bbb C})$.

We now consider two examples.  First, we consider the microscopic model
used in Ref.\ \onlinecite{BMSA} (and also Ref.\ \onlinecite{MBF}).
There, a distribution for the $t_n$ was assumed that was invariant under
unitary transformations of the chains,
\begin{equation}
  t_n = \exp(W_n),\ \ \ \
  \langle (W_{n})_{\mu \nu}^{\vphantom{*}}
  (W_{n})^{*}_{\rho \sigma} \rangle = \case{1}{2} {w^2 \beta}
  \delta_{\mu \rho} \delta_{\nu \sigma} .
 \label{eq:tW}
\end{equation} 
Here $W_n$ is a real (complex) matrix with independently and
identically Gaussian distributed elements for $\beta=1$ ($2$). In this
case, for small $w$, the distribution of the radial coordinates $x_j$
of the transfer matrix $M$ can be found explicitly. One obtains that
the diffusion rates $D_R$ and $D_S$ are equal, and the parameters $x_j$
obey a Fokker-Planck equation of the type (\ref{eq:DMPKchiral}) with
$D_{\rm ch} = w^2$ and $\eta=1$ \cite{BMSA,MBF}. We mention that the
most general Fokker-Planck equation (\ref{eq:DMPKchiral}), including
the dependence on $\eta$, can be derived from Eq.\ (\ref{eq:Schrod})
using $t_n = \exp(W_n)$ and a Gaussian distribution for the matrix
$W_n$ that involves correlations between the diagonal elements,
\begin{equation}
  \langle (W_{n})_{\mu \nu}^{\vphantom{*}}
  (W_{n})^{*}_{\rho \sigma} \rangle = \case{1}{2} {w^2 \beta}
  \left[\delta_{\mu \rho} \delta_{\nu \sigma} - 
  (1-\eta) N^{-1} \delta_{\mu \nu}
  \delta_{\rho \sigma} \right].
\end{equation}

As a second example, we consider the random flux model
\cite{Lee,Furusaki}. This model corresponds to a square lattice, where
each plaquette contains a random flux, see Fig.\ \ref{fig:1}b. The
hopping matrices $t_n$ have the form
\begin{equation}
  t_n = \left( \begin{array}{ccccc} 
    1 & e^{i \phi_{1,n}} & 0 & \ldots & 0 \\
    0 & 1 & e^{i \phi_{2,n}} & \ldots & 0 \\
    \vdots & \vdots &   & & \vdots \\
    0 & 0 &  \ldots & 1 & e^{i \phi_{N-1,n}} \\
    0 & 0 &  \ldots & 0 & 1 \end{array} \right).
\end{equation}
The phases $\phi_{j,n}$ are distributed such that the fluxes
$\Phi_{j,n} = \phi_{j,n+1} - \phi_{j,n}$ are independently
distributed.  In order to find the distribution of the radial
coordinates $x_j$ for the random flux model, we need to compute the
matrix $m$, and find the eigenvalues $\exp(2 x_j)$, $j=1,\ldots,N$, of $m
m^{\dagger}$, see Eq.\ (\ref{eq:solution}). We do not need to solve
this problem exactly: Each of the hopping matrices $t_n$ in the random
flux model has $\det t_n = 1$, which implies $\exp(2 N \bar x) = \det m
m^{\dagger} = 1$ for all lengths. Hence, the average radial coordinate
$\bar x$ does not diffuse, so that $D_R = 0$. We conclude that
$\eta = 0$ for the random flux model.

The reason why $D_R = 0$ or $\eta=0$ for the random flux model is that
$\det t_n = 1$ for all $n$. More generally, {\em all} random hopping
models which have $\det t_n = 1$ are described by
Eq.\ (\ref{eq:DMPKchiral}) with $\eta = 0$.  One example of such a
model is a random hopping model on a square lattice with only
randomness in the transverse hopping amplitudes.  
In the
general case, however, there will be both randomness in the transverse
and in the longitudinal hopping amplitudes, so that $\eta > 0$.

\begin{figure}
\epsfxsize=0.5\hsize
\hspace{0.25\hsize}
\epsffile{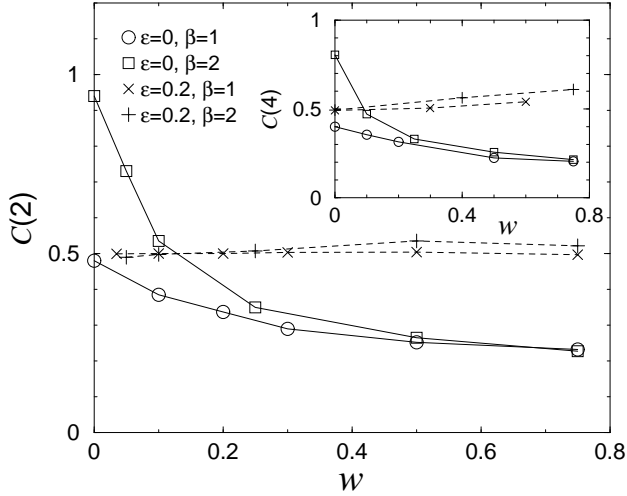}
\caption{\label{fig:3} The ratio $c(N) = -\lim_{L \to \infty} {\langle \ln g \rangle / \mbox{var}\, \ln g}$ as a function of the disorder strength
$w$ of the longitudinal hopping parameters, for fixed randomness of the
transversal hopping parameters. Statistical errors are smaller than the
sizes of the symbols.}
\end{figure}

To compare the Fokker-Planck equation (\ref{eq:DMPKchiral}) to
numerical simulations, we considered the quantity 
\begin{equation}
  c(N) \equiv 
  -\lim_{L \to \infty} {\langle \ln g \rangle \over \mbox{var}\, \ln g} =
  \left\{ \begin{array}{ll}
  \mbox{\large 
  ${\beta N/2 \over N + (1-2/\pi)(N-2+2\eta)}$},
  & \mbox{if $N$ is even}, \\ \\
  0, & \mbox{if $N$ is odd}. \end{array} \right. \label{eq:c}
\end{equation}
Here, we used Eqs.\ (\ref{eq:lnG}) and (\ref{eq:lnGodd}) for $\langle
\ln g \rangle$ and $\mbox{var}\, \ln g$. In the standard
case of on-site disorder, one has $c(N) = 1/2$ \cite{StoneReview}.
We have compared Eq.\ (\ref{eq:c}) to numerical simulations for a
square lattice of width $N$ and length $L$, attached to perfect
leads. For $\beta=1$, the transverse hopping amplitudes are taken 
from a uniform distribution in $[-0.2,0.2]$, whereas for $\beta=2$,
the transverse hopping amplitudes are complex numbers with amplitude
uniformly distributed in $[0,0.1]$ and a random phase.
In both cases the longitudinal 
hopping amplitudes are real numbers taken uniform in $[1-w,1+w]$. 
We have numerically computed the ratio $c(N)$ by taking an average over 
more than $2 \times 10^4$ samples. Results for $c(N)$ as a function of 
$w$ for $N=2$ and $N=4$ and for zero energy $\varepsilon=0$ are 
shown in Fig.\ \ref{fig:3}. 
In the presence of the chiral symmetry, i.e.\ for zero energy,
we expect that $c(N)$ depends on the details of the microscopic model 
(i.e., on the parameter $w$) through the parameter $\eta$. For $w=0$, 
we have $\eta=0$, and hence $c(2) = \beta/2$, $c(4) = \beta/(3 - 2/\pi)$. 
As the randomness $w$ in the longitudinal hopping amplitudes increases,
we expect an increase of $\eta$, and hence a decrease of $c(N)$,
which is confirmed by the numerical data shown in the figure. We have
also shown data for energy $\varepsilon=0.2$ where the chiral symmetry
is broken.\footnote{
        We have not shown the data points for $w=0$ and $N=2$ at finite
        energy, because in that case there is an extra reflection
        symmetry that is not taken into account in the standard
        DMPK equation.
} In this case we find $c(N) = 1/2$ for all $w$,
in agreement with the literature \cite{StoneReview}. [The slight 
increase of $c(N)$ with $w$ is attributed to a breakdown of the
weak disorder condition.]

Except for the case of the random flux model and the random-hopping
model with transverse random hopping, where $\eta=0$, we do not know
how to compute the parameter $\eta$ explicitly.  Numerical simulations
show that, typically, the parameter $\eta$ is of order unity. The
effect of a nonzero $\eta$ on the conductance distribution is most
pronounced for small $N$. For large $N$, the effect is small, because
only the radial coordinates near zero contribute to transport and the
additional correlations between these coordinates caused by a nonzero
$\eta$ decrease as $1/N$. Therefore, in the limit of large $N$, we
still expect a universal conductance distribution. Here, we would like
to remark that in Ref.\ \onlinecite{MBF}, the Fokker-Planck equation
(\ref{eq:DMPKchiral}) with $\eta=1$ was used to describe localization in
the random-flux model. Despite the fact that the random flux model
corresponds to $\eta=0$, excellent agreement was found between the
theory and numerical simulations performed for $N=15$ and up.

Before concluding, we would like to make two remarks. First, it is known that
universality of one-parameter scaling breaks down in (quasi-)
one dimensional disordered systems if there exist long-range
correlations in the disorder, for instance in periodic-on-average
systems \cite{Deych} or for random-hopping chains in the presence of a
staggering of the hopping parameter \cite{BMSA}, as is relevant
e.g.\ for narrow gap semiconductors \cite{OE} or charge-density wave
materials \cite{LRA}.  (In the latter case, the staggering of the
hopping parameter adds a drift term to the Brownian motion of $\bar x$,
whereas the Brownian motion of the relative positions of the $x_j$
remains unaffected \cite{BMSA}.) The nonuniversality that we discuss in
this paper is of quite a different origin: It follows directly from the
geometric structure of the transfer matrix group; no long-range
correlations in the disorder are involved, all the hopping amplitudes in
the examples we considered have independent distributions.
It should be noted that the appearance of a one-parameter family of
scaling equations also occurs in the case of random Dirac fermions in
two dimensions, where one finds a line of fixed points, rather than a
single fixed point \cite{Ludwig,Nersesyan,Mudry}.

Our second remark is of a more mathematical nature.
Altland and Zirnbauer \cite{AltlandZirnbauer,Zirnbauer} and Caselle
\cite{Caselle} have argued that Cartan's classification of all
symmetric spaces \cite{Helgason} offers a complete classification of
all possible random-matrix theories. These random-matrix theories
appear in triplets: distributions of eigenvalues of Hermitian matrices
(i.e., energy levels), distributions of eigenphases of unitary matrices
(i.e., scattering phase shifts), and a Fokker-Planck equation for the
radial eigenvalues of a transfer matrix.  It is the latter kind of
random-matrix theories that we have considered here. While Cartan's
classification is complete for all semi-simple transfer matrix groups,
it does not take into account the phenomenon that we have presented in
this paper:  that for some disordered systems the transfer matrix group
is not semi-simple, so that they cannot be represented by a single
element from Cartan's table.

To summarize, we have considered the Fokker-Planck equation for the
transmission eigenvalues of a quantum wire with off-diagonal (hopping)
disorder from a geometric point of view. Under the same assumption
of weak disorder that leads to the universal DMPK equation
in the standard case of diagonal (on-site) disorder \cite{Dorokhov,MPK},
we have found that
the Fokker-Planck equation for the transmission eigenvalues of a quantum wire 
with off-diagonal disorder contains an extra parameter that 
depends on the microscopic details of the disorder. The existence of 
this extra parameter leads to a {\em nonuniversality} of transport 
properties in random-hopping chains,
which is most prominent if the number $N$ of coupled chains is small.

We would like to thank D.\ S.\ Fisher and
B.\ I.\ Halperin for discussions.  PWB acknowledges support by the NSF
under grant nos.\ DMR 94-16910, DMR 96-30064, and DMR 97-14725.  CM
acknowledges a fellowship from the Swiss Nationalfonds. The numerical
computations were performed at the Yukawa Institute Computer Facility.

\end{document}